\RequirePackage{pdf14}

\documentclass[conference]{IEEEtran}
\IEEEoverridecommandlockouts
\usepackage{balance}

\usepackage{graphicx}\graphicspath{{./images/}}
\usepackage{color, xcolor, makecell}
\def\BibTeX{{\rm B\kern-.05em{\sc i\kern-.025em b}\kern-.08em
    T\kern-.1667em\lower.7ex\hbox{E}\kern-.125emX}}
		
\ifCLASSOPTIONcompsoc
\usepackage[caption=false, font=normalsize, labelfont=sf, textfont=sf]{subfig}
\else
\usepackage[caption=false, font=footnotesize]{subfig}
\fi
\usepackage{latexsym, bm}
\usepackage{nohyperref}	
\hypersetup{colorlinks, pdfa=true, breaklinks, citecolor=black, urlcolor=black, linkcolor=black}
\newcommand*{\email}[1]{\normalsize\texttt{\href{mailto:#1}{#1}}\par}

\usepackage{cite}
\usepackage{mathtools, mdwmath, mdwtab, multirow, amssymb, amsmath, array, ragged2e}

\usepackage{threeparttable, booktabs, url, eqparbox, xspace, setspace}
\makeatletter\let\MPtrue\@minipagetrue\makeatother

\usepackage{algorithm, algorithmicx, algpseudocode}
\algnewcommand{\LineComment}[1]{\Statex \hfill\hskip\ALG@thistlm \textcolor[rgb]{0,0,0.55}{\(\triangleright\)~#1}}
\algrenewcommand\alglinenumber[1]{{\sffamily\footnotesize#1}}
\usepackage{etoolbox}\AtBeginEnvironment{algorithmic}{\small}

\algnewcommand{\Initialize}[1]{\State \textbf{Initialize:}\\%
      \hspace*{\algorithmicindent}\MPtrue\raggedright #1}

\algnewcommand\algorithmicswitch{\textbf{switch}}
\algnewcommand\algorithmiccase{\textbf{case}}
\algnewcommand\algorithmicassert{\texttt{assert}}
\algnewcommand\Assert[1]{\State \algorithmicassert(#1)}%
\algdef{SE}[SWITCH]{Switch}{EndSwitch}[1]{\algorithmicswitch\ #1\ \algorithmicdo}{\algorithmicend\ \algorithmicswitch}%
\algdef{SE}[CASE]{Case}{EndCase}[1]{\algorithmiccase\ #1}{\algorithmicend\ \algorithmiccase}%
\algtext*{EndSwitch}%
\algtext*{EndCase}%
\makeatletter
\newcommand\fs@betterruled{%
	\def\@fs@cfont{\bfseries}\let\@fs@capt\floatc@ruled
	\def\@fs@pre{\vspace*{8pt}\hrule height.8pt depth0pt \kern2pt}%
	\def\@fs@post{\kern2pt\hrule\relax}%
	\def\@fs@mid{\kern2pt\hrule\kern2pt}%
	\let\@fs@iftopcapt\iftrue}
\floatstyle{betterruled}
\restylefloat{algorithm}
\makeatother

\usepackage{xpatch}
\makeatletter
\xpatchcmd{\algorithmic}{\itemsep\z@}{\itemsep=0.4ex plus1pt}{}{}
\makeatother


\usepackage{hyphenat}

\makeatletter
\DeclareRobustCommand\onedot{\futurelet\@let@token\@onedot}
\def\@onedot{\ifx\@let@token.\else.\null\fi\xspace}
\def\eg{\emph{e.g}\onedot} 
\def\ie{\emph{i.e}\onedot}

\def\etal{\emph{et al}\onedot}
\makeatother

\usepackage{xpatch}

\ifCLASSOPTIONcompsoc
 \usepackage[caption=false,font=normalsize,labelfont=sf,textfont=sf]{subfig}
\else
 \usepackage[caption=false,font=footnotesize]{subfig}
\fi

\def\BibTeX{{\rm B\kern-.05em{\sc i\kern-.025em b}\kern-.08em
    T\kern-.1667em\lower.7ex\hbox{E}\kern-.125emX}}
    
    \makeatletter
\newcommand{\linebreakand}{%
  \end{@IEEEauthorhalign}
  \hfill\mbox{}\par
  \mbox{}\hfill\begin{@IEEEauthorhalign}
}

\def\ps@IEEEtitlepagestyle{%
  \def\@oddfoot{\mycopyrightnotice}%
  \def\@evenfoot{}%
}
\def\mycopyrightnotice{%
  {\footnotesize 978-1-7281-8402-9/21/\$31.00 \copyright 2021 IEEE\hfill}
  \gdef\mycopyrightnotice{}
}

\@ifundefined{showcaptionsetup}{}{
 \PassOptionsToPackage{caption=false}{subfig}}
\usepackage{subfig}
\makeatother

\usepackage{eso-pic}
\newcommand\AtPageUpperMyleft[1]{\AtPageUpperLeft{%
 \put(\LenToUnit{0.08\paperwidth},\LenToUnit{-1cm}){%
     \parbox{0.9\textwidth}{\raggedright\fontsize{9}{11}\selectfont #1}}%
 }}%

\begin{document}

\title{A Combined PCA-MLP Network for Early\\ Breast Cancer Detection}


\author{
    \IEEEauthorblockN{%
    Md. Wahiduzzaman Khan Arnob\IEEEauthorrefmark{1}, Arunima Dey Pooja\IEEEauthorrefmark{2} and Md. Saif Hassan Onim\IEEEauthorrefmark{3}
                    }

    \IEEEauthorblockA{%
        Department of Electrical, Electronic and Communication Engineering,\\
        Military Institute of Science and Technology (MIST) Dhaka-1216, Bangladesh}

\IEEEauthorrefmark{1}\email{arnobk511@gmail.com}, \IEEEauthorrefmark{2}\email{arunima.pooja99@gmail.com}, \IEEEauthorrefmark{3}\email{saif@eece.mist.ac.bd}
}

\maketitle

\begin{abstract}
Breast cancer is the second most responsible for all cancer types and has been the cause of numerous deaths over the years, especially among women. Any improvisation of the existing diagnosis system for the detection of cancer can contribute to minimizing the death ratio. Moreover, cancer detection at an early stage has recently been a prime research area in the scientific community to enhance the survival rate. Proper choice of machine learning tools can ensure early-stage prognosis with high accuracy. In this paper, we have studied different machine learning algorithms to detect whether a patient is likely to face breast cancer or not. Due to the implicit behavior of early-stage features, we have implemented a multilayer perception model with the integration of PCA and suggested it to be more viable than other detection algorithms. Our 4 layers MLP-PCA network has obtained the best accuracy of 100\% with a mean of 90.48\% accuracy on the BCCD dataset.\end{abstract}


\section{Introduction}

 
 Breast cancer is recognised as one of the most widespread diseases and one of the leading causes of death amongst women globally. It is reported as the most common cancer amid women worldwide by World Health Organization (WHO). In 2020, 2.3 million women were diagnosed and 0.685 million died around the world\cite{whoBreastCancer}. Breast cancer originates from the breast tissue where the cells start to grow abnormally. The growth of abnormal cells form a lump called tumor. A tumor can be benign, very similar to any other normal cells in appearance and non cancerous or malignant, highly cancerous as it grows irresistibly and spreads in other parts of the body. Initially the affected cells do not show any symptoms. But over time the cancer cells progress effecting the surrounding tissues and organs. Eventually becoming life threatening. But early prognosis of this life taking cancer can boost the medication procedure significantly with a reduced mortality rate of 25\% \cite{businesswireGRAILAnnounces,gvamichava2012cancer, khairunnahar2019classification}.\\
 
 Early detection of breast cancer may prevent death but it is difficult as the symptoms do not reveal at the beginning. Mammogram \cite{nag2017identifying} and biopsy\cite{lopez2012digital} have been seen as promising screening processes used in the breast tumors detection. Mammography procedure enables the identification of tumorous cell at a primary stage even when it is still not diagnosed by the clinical examination or has not yet started to show symptoms \cite{singh2020breast, amrane2018breast}.
 It is difficult to predict the normal tissue and malignant tissue in case of low contrast mammograms\cite{sundaram2011histogram}. Again, identifying benign and malignant tumours cells through biopsy is a very time consuming and complex process\cite{chekkoury2012automated}. The doctors often misdiagnosis due to lower attributes differences between cancerous and normal tissues and irregularity of cells' size\cite{mehdy2017artificial, pollanen2014computer}. The accurate examination of breast cancer and its classification into benign or malignant categories is a concern of much exploration. The consequences of the failure in early detection leads to the cancer cells reaching a stage where it becomes both difficult and expensive to cure fully.
 Machine learning (ML) has been considered as a promising cancer detection approach due to its ability to distinguish attributes from complex data sets\cite{khan2019novel}. It acts as a rescuer by developing new features from the existing data of the low contrast mammography images. Recent literature shows an increasing usage of different ML techniques cultivated in this field.\\
 
 Ghosh~\etal~\cite{ghosh2014} made a comparative study using Support vector machine(SVM) and Multilayer perception (MLP) with back propagation neural network based on the original Wisconsin Breast Cancer Dataset (WBCD)~\cite{mangasarian1990}. They opted for SVM as it gave an accuracy of 96.71\% compared to 95.71\% for MLP.
 Rana~\etal~\cite{rana2015} did a performance analysis using popular ML algorithms \ie SVM, Logistic regression (LR), k-Nearest Neighbors(KNN) and Naive Bayes. The best accuracy achieved was 95.68\% for KNN, but SVM with Gaussian kernel was chosen as the suitable technique due to its better performance.
 Khuriwal ~\etal~\cite{khuriwal2018icacccn} applied deep learning with convolutional neural network on the Wisconsin breast cancer dataset (diagnostic)\cite{Dua:2019}. They gave a fairly better result than previous literature. Their obtained accuracy was 99.67\% using only 12 features dataset.
 kadam ~\etal~\cite{kadam2019} proposed a feature ensemble learning model on the similar dataset that showed accuracy of 98.60\%. The model was based on sparse auto-encoders and softmax regression exhibiting better performance than the existing ML algorithm. Most of these approaches were applied on dataset extracted from cells that have already exhibited cancerous symptoms.\\ 
 

A detection process with biomarkers has been introduced recently on the Breast Cancer Coimbra Dataset (BCCD)\cite{patricio2018}. Here data is taken from features like age, body mass index (BMI) and proteins found in blood \eg \textit{glucose, resistin, insulin, leptin, HOMA, adiponectin, MCP-1}. These are considered as potential bio marker for early stage breast cancer prognosis before any clinical diagnosis.
Patricio~\etal~\cite{patricio2018} evaluated a performance comparison of LR, Random forest (RF) and SVM where they suggested attribute set containing \textit{resistin, glucose,} BMI and age as suitable biomarkers for practical screening test.
Li~\etal~\cite{li2018} did a comparative study with 5 ML algorithms \ie Decision Tree (DT), SVM, RF, LR and Neural Network (NN) where RF displayed best accuracy of 74.3\% in predicting the presence of tumor. A similar research was done by Ghani ~\etal~\cite{ghani2019} with different classification models \ie KNN, DT and Naive Bayes. Their work showed a better performance than the previous stated work with 80.00\% highest accuracy and 77.14\% average accuracy. Kusuma~\etal~\cite{kusuma2020} adopted a different approach on the same dataset with back propagation neural network based on Neldermead classification. Their best result was found to be 76.5\% and in average 73.4\%.\\

Owing to much work in classifying cancerous cell before clinical diagnosis there is a wide gap in the accuracy which needs to be addressed adequately. In our work, we focus on closing the gap between highest accuracy and computational complexity without any significant loss of execution time. We propose our algorithm on the well known BCCD with 15 fold cross validation and minor preprocessing.

\section{Methodology}
\subsection{Multilayer Perceptron}

\begin{figure}[tbhp]
\centering
\includegraphics[width=0.7\linewidth]{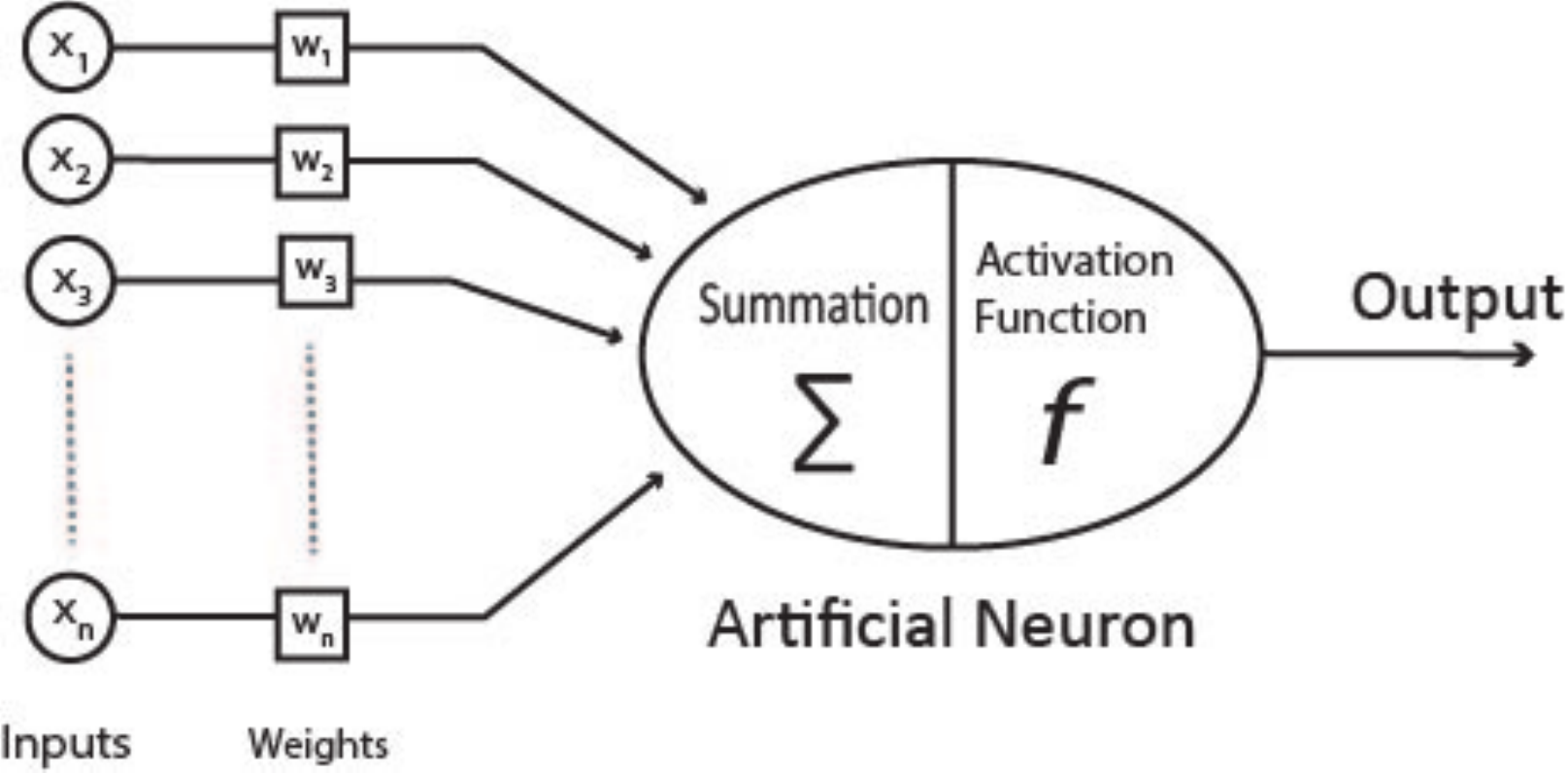}
\caption{Artificial Neuron}
\label{Artificial Neuron}
\end{figure}

A multilayer perceptron (MLP) is a combination of neurons called perceptions, where each of the neuron is a representation of a function. It is represented as a node shown in Fig.\ref{Artificial Neuron}

\subsection{Principal Component Analysis}

\begin{figure}[tbhp]
\centering
\includegraphics[width=0.9\linewidth]{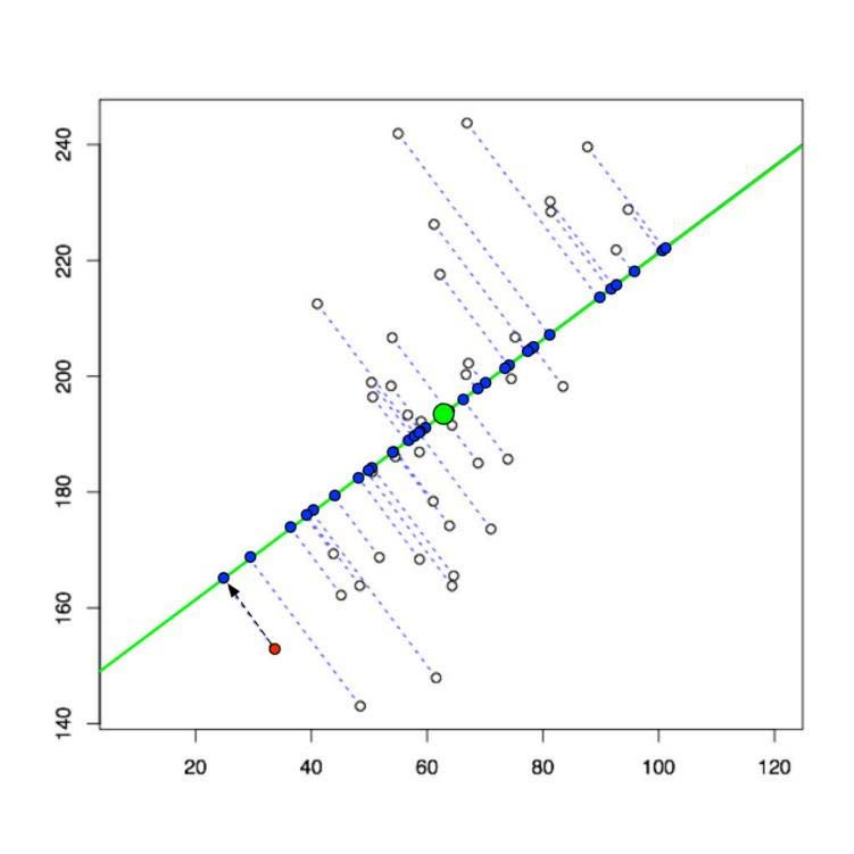}
\caption{Dimensionality reduction with data projection}
\label{Dimensionality reduction with data projection}
\end{figure}

Principal component analysis (PCA) is a dimensioanlity reduction method where a multivariate data table is presented with smaller group of variables with the help of data projection. A graphical representation in shown in Fig. \ref{Dimensionality reduction with data projection}.

\subsection{Activation Function}
\begin{figure}[tbhp]
\centering
\includegraphics[width=0.7\linewidth]{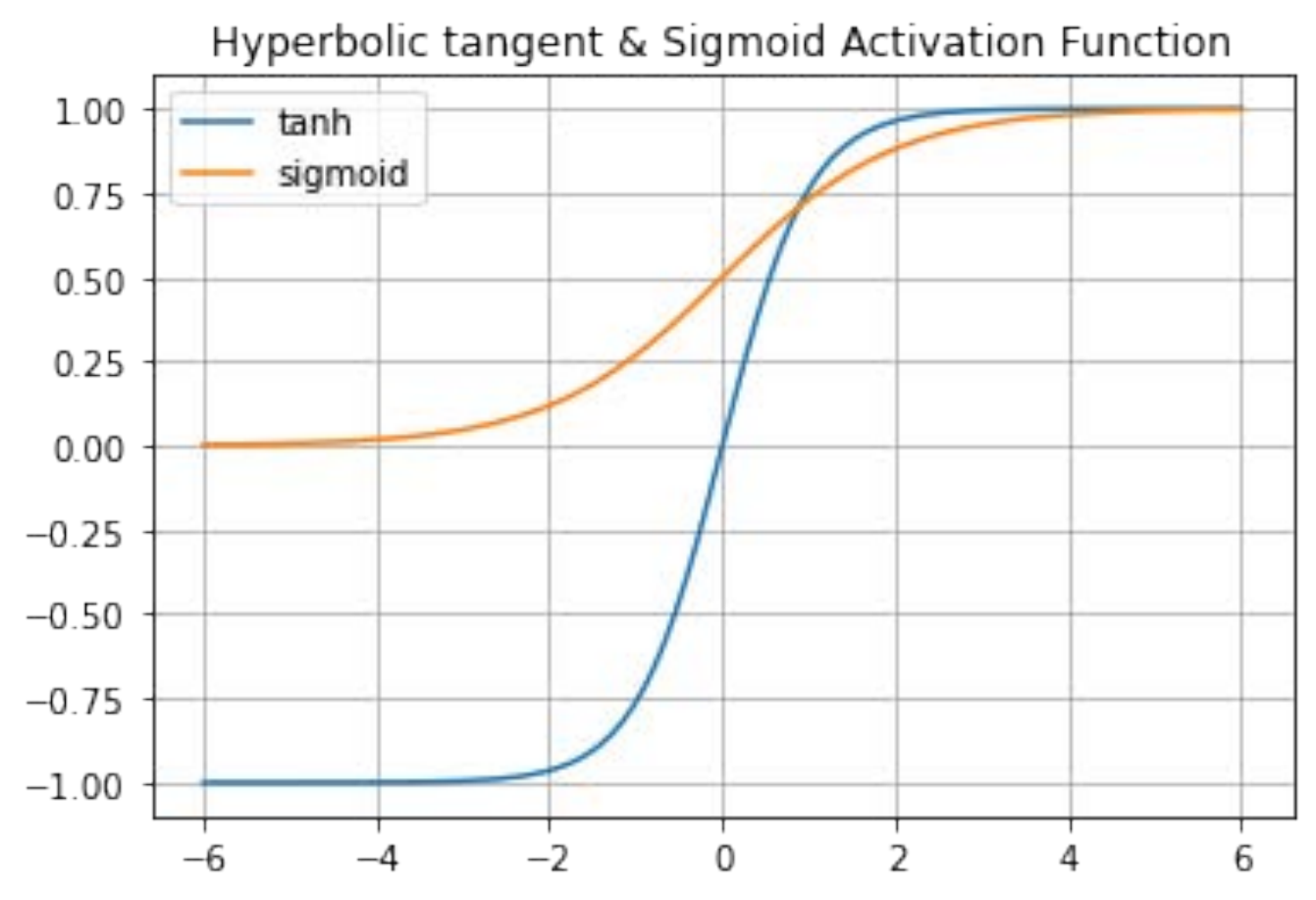}
\caption{Hyperbolic tangent and sigmoid activation function}
\label{Hyperbolic tangent and sigmoid activation function}
\end{figure}

It is a function that determines the outcome of a neural network model by controlling the neuron whether it should be activated or not. Two activation functions \ie hyperbolic tangent function and \textit{sigmoid} function used in or work are shown in Fig.\ref{Hyperbolic tangent and sigmoid activation function}. Equation \eqref{eq:activation function} denotes their mathematical representations.

\begin{equation}
\label{eq:activation function}
tanh(x)={e^{x}-e^{-x}}/{e^{x}+e^{-x}}, 
sig(x)={1}/{1+e^{-x}}
\end{equation}

\subsection{Dataset Description}

BCCD contains 10 predictors with 9 real valued quantitative attributes and 1 label classifier indicating the presence of breast cancerous cells. The features are shown in Table \ref{tab-features}.

\begin{table}[tbhp]
	\centering
	\renewcommand{\arraystretch}{1.1}
	\setlength{\tabcolsep}{8pt}
	\caption{Features in Dataset}
	\label{tab-features}
	\resizebox{.6\columnwidth}{!}{
		    \begin{tabular}{l|c|c}
    \toprule
        Parameter & Unit & Datatype \\
        \midrule
        Age  & years  & int64  \\
        BMI & kg/m2 & float64 \\
        Glucose & mg/dL & int64 \\
        Insulin & µU/mL & float64 \\
        HOMA & ng/mL & float64 \\
        Leptin & ng/mL & float64 \\
        Adiponectin & µg/mL & float64 \\
        Resistin & ng/mL & float64 \\
        MCP.1 & pg/mL & float64 \\
        \midrule
        Classification & \makecell[c]{1=Healthy\\ 2=Patients}  & int64 \\
    \bottomrule
    \end{tabular}
    }
\end{table}

\begin{figure}[tbhp]
\centering
\includegraphics[width=0.8\linewidth]{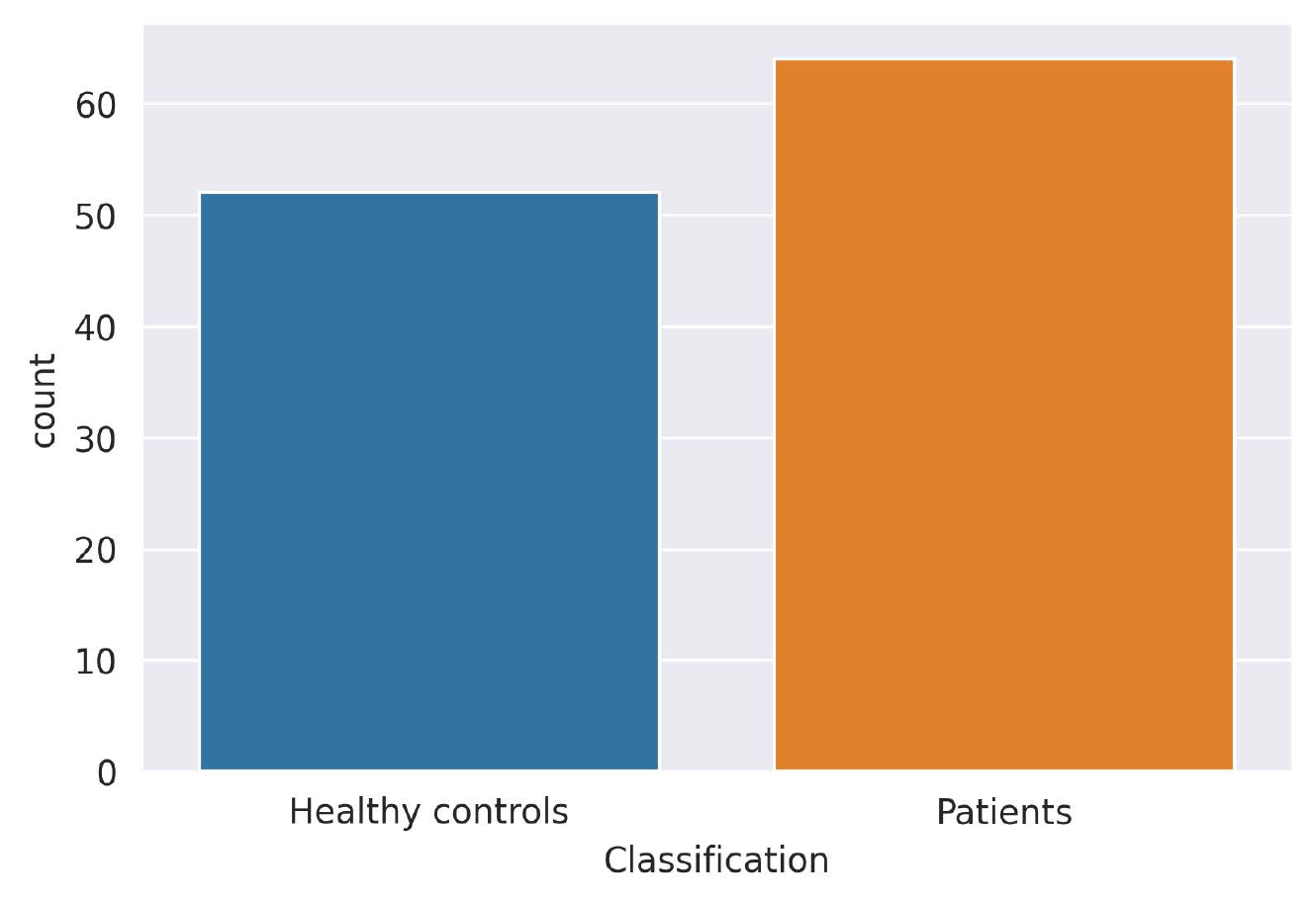}
\caption{Number of Healthy Controls and Breast cancer patients}
\label{Number of Healthy Controls and Breast cancer patients}
\end{figure}

A total number of 116 instances are present in the dataset collected from the routine blood test of 64 breast cancer patients (52.17\%) and 52 healthy people (44.8\%). As the dataset is fairly balanced, no over sampling process is required here. Fig. \ref{Number of Healthy Controls and Breast cancer patients} displays the bar diagram of total cases of breast cancer patients and healthy controls.

\subsection{Dataset Preprocessing}
\subsubsection{Data Transformation}

Due to a wide range of data, standard scaling was done on the dataset through which the attributes were transformed to a range from $-1.0$ to $+1.0$. It reduces the variation of the data and thus making the calculation easier. The formula for standard scaling used is shown in Equation \eqref{eq:standard scaling}.

\begin{equation}
\label{eq:standard scaling}
z = \frac{\left( x - \mu \right)}{\sigma}
\end{equation}

Here, x = Original feature values of the dataset, $\mu$ = Mean of the feature values, $\sigma$ = Standard deviation of the feature values and z = Standard scaled feature values. For example, mean value for the feature insulin is 10.01208 and the standard deviation is 10.0242. Now if the amount of insulin in a patient is 2.707 (µU/mL) then the scaled value will be -0.7287.\\

\subsubsection{Dimensionality Reduction}
PCA is applied for dimensionality reduction. The theme of PCA is to find such a line, surface or space upon which the data is projected with least error. To reduce the dataset from n-dimension to k-dimension following steps are needed.
\begin{enumerate}

\item Firstly the co-variance matrix is computed through the equation :

\begin{equation}\label{covarience}
\Sigma = \frac{1}{m}\sum_{i=1}^{n}(x^{(i)})(x^{(i)})^{T}
\end{equation}

\item Then the \textit{eigen vectors} of matrix $\Sigma$ is computed. For Example, if n-dimension to k-dimension reduction is required then k number of vectors $(u^{(1)}), (u^{(2)}),...., (u^{(k)})$ are calculated from singular value decomposition.

\item Finally the data points created by the vectors $(u^{(1)}), (u^{(2)}),...., (u^{(k)})$ are projected on the surface.

\end{enumerate}

Thus the dimension of the feature matrix is reduced from n-dimension to k-dimension.

\subsubsection{Folding and Shuffling}
For Cross validation data is folded 15 times where in each fold, one of the sets is used as test dataset. The original dataset is sorted having 1st 52 examples as class 1 or healthy controls and last 64 example as class 2 or breast cancer patients. Directly feeding this data to the model will create a class imbalance resulting poor performance. Therefore random shuffling of data is performed to ensure data diversity.

\subsection{Proposed Method}

\begin{figure*}[tbhp]
\centering
\includegraphics[width=1.8\columnwidth, height=2.5in]{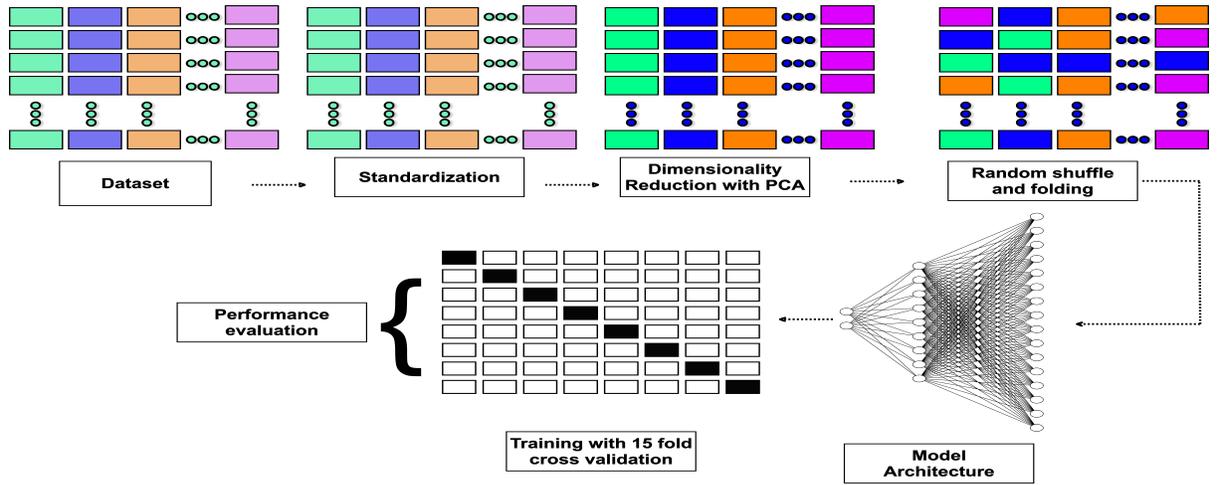}
\caption{Flow diagram of the Process}
\label{Flow diagram of the Process}
\end{figure*}

We worked on BCCD that consists biomarkers from blood analysis. We started by standardising the data to make the mean value 0 and standard variation 1 for each feature. After that we applied PCA for dimensionality reduction and trialed our model for 2 to 9 each dimension. Before the trial we shuffled and folded the data for 15 times for cross validation. Finally we built our model with a MLP and compiled it. The proposed model is displayed in Fig. \ref{Flow diagram of the Process} as a flow diagram.

\subsection{Neural Network Model and Training}

\begin{figure*}[tbhp]
\centering
\includegraphics[width=1.8\columnwidth, height = 2in]{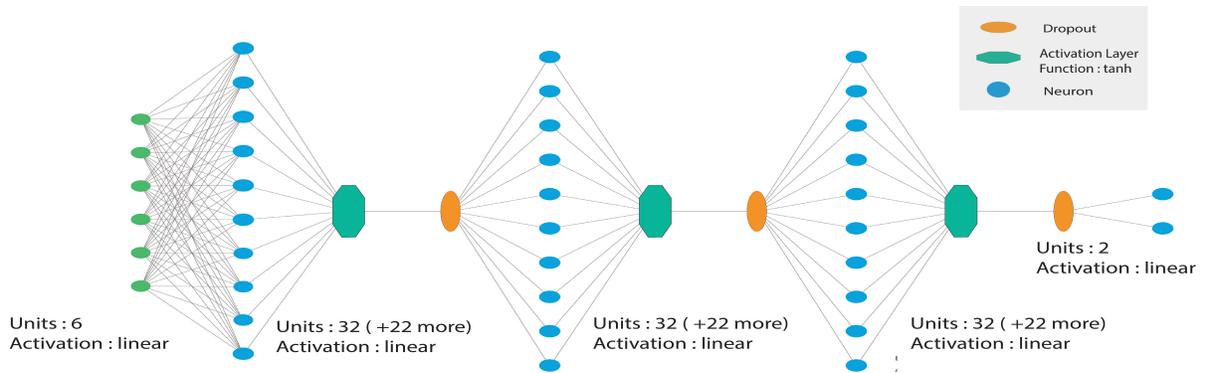}
\caption{Model architecture}
\label{Model architecture}
\end{figure*}

A four layer perceptron model is built where the number of units in the input layer is same as the feature numbers of input data. Here we have trialed the input layer taking 2 to 9(all) features by applying PCA. Each hidden layer has 32 units making the model cylindrical shaped. We have used \textit{tanh} activation function in input layer and hidden layers. As we are doing binary classification, \textit{sigmoid} activation function is used at the output layer. The MLP model is shown in Fig.\ref{Model architecture}. We fit the model for each folded training data and at the same time evaluated them with validation data. At the end we calculated the mean and standard deviation from the evaluation metrics gained for each fold.\\

\begin{table}[!th]
	\centering
	\renewcommand{\arraystretch}{1}
	\setlength{\tabcolsep}{4pt}
	\caption{Comparison of hyper-parameters}
	\label{tab-hyperparameter}
	\resizebox{\columnwidth}{!}{
	\begin{tabular}{lccc}
\toprule
Authors &  Method & \makecell[c]{Parameters}\\
\midrule

Ghosh~\etal~\cite{ghosh2014} &\makecell[c]{MLP \& SVM} & \makecell[l]{MLP BPN :\\Learning rate = 0.3, Momentum = 0.2\\ Activation = tanh\\SVM:\\ Kernel = poly, C (cache size) = 250007 \\ E (exponent) = 1.0, Epochs = 500}\\
\midrule

Rana~\etal~\cite{rana2015} &\makecell[c]{SVM, LR, KNN \\ Naive Bayes} & \makecell[l]{Activation = sigmoid}\\
\midrule

Khuriwal~\etal~\cite{khuriwal2018icacccn} &\makecell[c]{CNN} & \makecell[l]{Activation = sigmoid, Epochs=20 \\ Number of Neurons = 12}\\
\midrule

Kadam~\etal~\cite{kadam2019}  & \makecell[c]{FE-SSAE-SM} & \makecell[l]{Epochs : 50 - 600, Regularization, $\lambda = 0.005 - 8$\\ Sparsity Proportion, $\rho = 0.01 - 1$ \\Sparsity regularization, $\beta = 1 - 10$}\\
\midrule

Patricio~\etal~\cite{patricio2018} &\makecell[c]{LR, RF\\ \& SVM} & \makecell[l]{Random splits (MCCV) = 500, Epochs = 100}\\
\midrule

Li~\etal~\cite{li2018} &\makecell[l]{DT, RF, SVM\\ LR \& NN} & \makecell[l]{Not mentioned}\\
\midrule

Ghani~\etal~\cite{ghani2019} & \makecell[c]{K-NN, DT\\ Naive Bayes} & \makecell[l]{Not mentioned}\\
\midrule

Kusuma~\etal~\cite{kusuma2020} & \makecell[c]{NM-BPNN} & \makecell[l]{Nelder Mead operation parameters\\ expansion : $\chi = 2$, reflection : $\rho = 1$ \\contraction : $\gamma = \frac{1}{2}$, shrinkage : $\alpha = \frac{1}{2}$}\\
\midrule
\textbf{Ours} & \makecell[c]{\bf PCA-MLP} & \makecell[l]{Optimization algorithm = Adam\\Learning rate = $0.00001$, $\beta 1 = 0.90$, $\beta 2 = 0.99$\\Dropout = $0.45$, Batch size = $15$, Epochs=3000} \\
\bottomrule
\end{tabular}
}
\end{table}

\begin{figure}[h]
\centering
\includegraphics[width=0.9\linewidth]{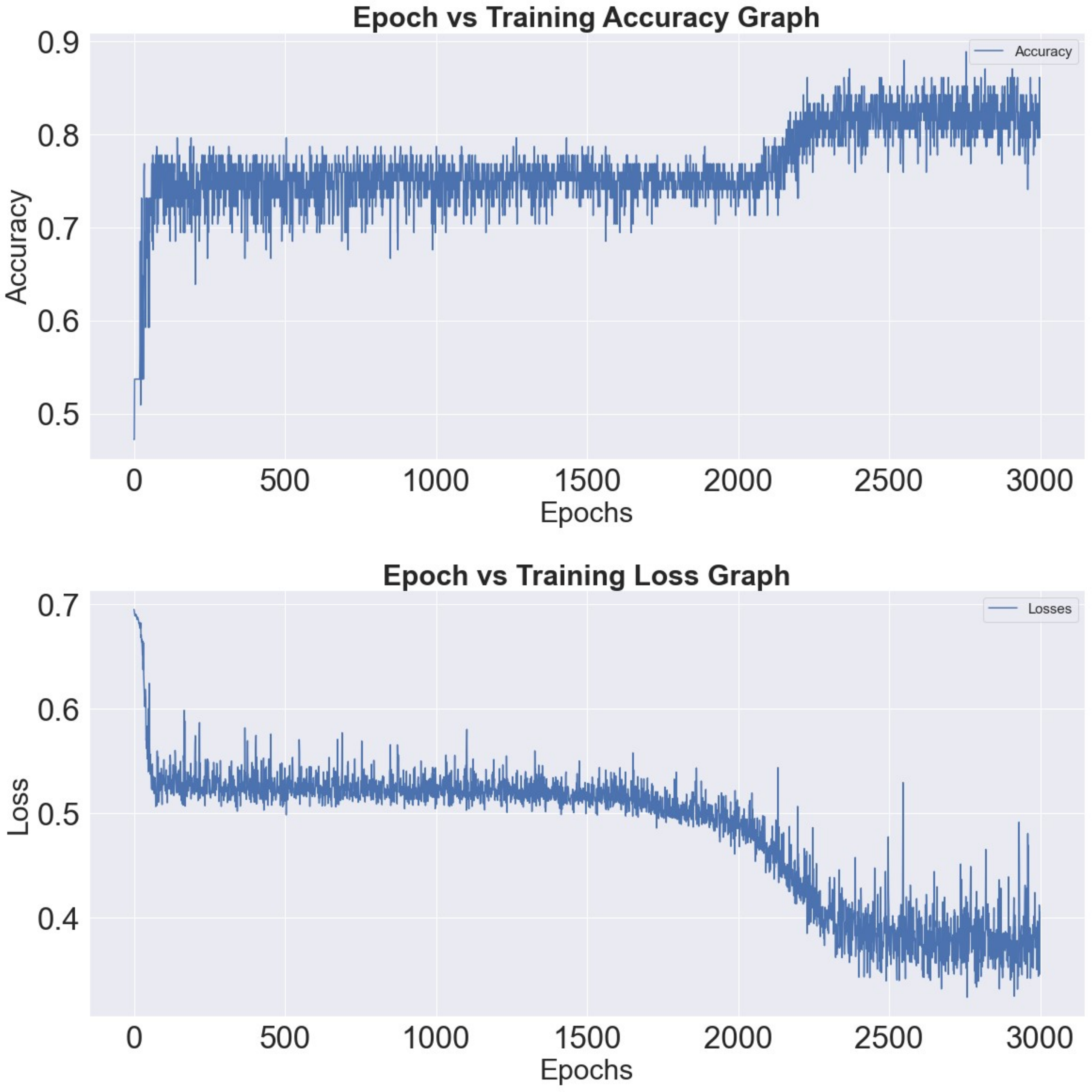}
\caption{Accuracy and loss curve}
\label{Accuracy and loss curve}
\end{figure}

To avoid overfitting we have applied 45\% drop out regularization. The proposed model is trained with 15 fold cross validation and training batch size is 15. While fitting the training data we applied callback to stored the weights that gives the best result for each validation. Finally, we evaluated their performance metrics to achieve mean value.\\

A comparative lists of all hyper-parameters used in our work and recent prominent methods are mentioned in Table \ref{tab-hyperparameter}.

\section{Result Analysis}
\begin{table*}[tbhp]
	\centering
	\renewcommand{\arraystretch}{.9}
	\setlength{\tabcolsep}{8pt}
	\caption{Comparison with Recent Works}
	\label{tab-comparison}
	\resizebox{1.8\columnwidth}{!}{
		    \begin{tabular}{lccccc}
\toprule
Year & Authors & Dataset & Detection & Mathod & \makecell[c]{Metrics\\ (Accuracy)}\\
\midrule
2014  & Ghosh~\etal~\cite{ghosh2014} & \makecell[l]{Breast Cancer\\Wisconsin(Original)} & Late & \makecell[c]{MLP \& SVM} & \makecell[l]{$Best: 96.71\%$ \\ $Avg:N/A $}  \\
\midrule
2015 & Rana~\etal~\cite{rana2015breast} & \makecell[l]{Breast Cancer\\Wisconsin} & Late & \makecell[c]{SVM, LR, KNN \\ Naive Bayes} & \makecell[l]{$Best: 95.68\%$ \\ $Avg:N/A $} \\
\midrule
2018 & Khuriwal~\etal~\cite{khuriwal2018icacccn} & \makecell[l]{Breast Cancer\\Wisconsin(Diagnostic)} & Late & \makecell[c]{LR \& NN} & \makecell[l]{$Best: 99.67\%$ \\ $Avg: N/A$}\\
\midrule
2019 & Kadam~\etal~\cite{kadam2019} & \makecell[l]{Breast Cancer\\Wisconsin(Diagnostic)} & Late &\makecell[c]{FE-SSAE-SM} & \makecell[l]{$Best: 98.60\%$ \\ $Avg:N/A$}\\
\midrule
2018 & Patricio~\etal~\cite{patricio2018} & \makecell[l]{Breast Cancer\\Coimbra Dataset} & Early &\makecell[c]{LR, RF\\ \& SVM} & \makecell[l]{$Best:N/A $ \\ $Avg:N/A $}\\
\midrule
2018 & Li~\etal~\cite{li2018} & \makecell[l]{Breast Cancer\\Coimbra Dataset} & Early &\makecell[c]{DT, RF, SVM\\ LR \& NN} & \makecell[l]{$Best: 74.3\%$ \\ $Avg:N/A$}\\
\midrule
2019 & Ghani~\etal~\cite{ghani2019} & \makecell[l]{Breast Cancer\\Coimbra Dataset} & Early &\makecell[c]{K-NN, DT\\ Naive Bayes} & \makecell[l]{$Best: 80.00\%$ \\ $Avg: 77.14\%$}\\
\midrule
2020 & Kusuma~\etal~\cite{kusuma2020} & \makecell[l]{Breast Cancer\\Coimbra Dataset} & Early &\makecell[c]{NM-BPNN} & \makecell[l]{$Best: 76.5217\%$ \\ $Avg: 73.3844\%$}\\
\midrule
\textbf{2022} & \textbf{Ours} & \makecell[l]{\bf Breast Cancer\\ \bf Coimbra Dataset} & \textbf{Early} & \makecell[c]{\bf MLP} & \makecell[l]{$\mathbf{Best: 100.00.\%}$ \\ $\mathbf{Avg: 90.48\%}$} \\
\bottomrule
\end{tabular}
}
\end{table*}


Fig.\ref{Accuracy and loss curve} displays the accuracy and loss curve where after 3000 epochs, both accuracy and loss have been saturated. Therefore the parameter for epoch is chosen as 3000. With this curve we have set the values of hyper-parameter in table \ref{tab-hyperparameter} through test and trial.

A comparison is shown in table \ref{tab-comparison}. It is seen that late detection accuracy is highest for kadam~\etal~\cite{kadam2019}. Again other researches regarding early detection could not show any reliable results. With our work, we have achieved best accuracy where our models predicts all samples correctly averaging more than 90\% accuracy which is the highest among the previous literature.

\section{Conclusions}
\label{conclusions}\thispagestyle{empty}
In support of developing an early stage prognosis for breast cancer, we have introduced a 4 layer MLP model incorporating with PCA. The proposed model has been investigated on the BCCD biomarkers which is comparatively a new approach for detecting cancerous cells. As the dataset is quite small it has been quite a challenge to obtain a reliable accuracy and validation result. But our combined PCA - MLP network has proven to be promising with very satisfactory results for the early state detection. It outperforms recent literature by 5\% to 17\% in mean and best detection accuracy.\looseness -1

\balance

\bibliographystyle{IEEEtran}
\bibliography{ref}
\end{document}